\documentclass[preprint]{aastex}

\usepackage{epstopdf} \usepackage{subfigure}

   \def\gray{$\gamma$-ray}

     \def\etal{et     al.~}
\def\today{\ifcase\month\or  January\or February\or  March\or April\or
May\or June\or  July\or August\or September\or  October\or November\or
December\fi
\space\number\day, \number\year}
\begin{document}
\title{Is the Universe More  Transparent to Gamma Rays than Previously
Thought?}

\author{Floyd W.  Stecker}  \affil{Astrophysics Science Division, NASA
Goddard  Space Flight  Center  \\ Greenbelt,  MD  20771, U.S.A.\\  \it
Floyd.W.Stecker@nasa.gov\rm}

\smallskip

\author{Sean T. Scully} \affil{Department of Physics, \\ James Madison
   University,     Harrisonburg,     VA     22807,    U.S.A.\\     \it
   scullyst@jmu.edu\rm}
\slugcomment{Submitted to {\it The Astrophysical Journal Letters}}
%

\begin{abstract} 

The  MAGIC collaboration has  recently reported  the detection  of the
strong  \gray\ blazar 3C279  during a  1-2 day  flare. They  have used
their spectral observations to draw conclusions regarding upper limits
on  the  opacity of  the  Universe  to high  energy  \gray  s and,  by
implication, upper limits on the extragalactic mid-infrared background
radiation. In this paper we examine the effect of \gray ~absorption by
the  extragalactic infrared  radiation on  intrinsic spectra  for this
blazar  and  compare  our  results  with  the  observational  data  on
3C279. We  find agreement with  our previous results, contrary  to the
recent  assertion  of  the  MAGIC  group that  the  Universe  is  more
transparent to  \gray s than our calculations  indicate.  Our analysis
indicates that  in the energy range  between $\sim$ 80  and $\sim$ 500
GeV, 3C279  has a  best-fit intrinsic spectrum  with a  spectral index
$\sim$ 1.78 using  our fast evolution model and  $\sim$ 2.19 using our
baseline model.   However, we also  find that spectral indices  in the
range of 1.0  to 3.0 are almost as equally acceptable  as the best fit
spectral indices.  Assuming the same intrinsic spectral index for this
flare as for the 1991 flare  from 3C279 observed by EGRET, viz., 2.02,
which lies  between our best fit  indeces, we estimate  that the MAGIC
flare  was $\sim$3  times brighter  than the  EGRET flare  observed 15
years earlier.

\end{abstract}
\section{INTRODUCTION}
 \label{sec:intro}

It has  long been  recognized that by  studying the spectra  of strong
extragalactic  \gray\ sources  one  can obtain  information about  the
density  and  energy  spectra  of intergalactic  photon  fields.   The
luminous blazar 3C279 was discovered  by the EGRET detector aboard the
Compton Gamma  Ray Observatory  to be a  strong flaring  \gray\ source
(Hartman  et al. 1992).   Shortly after  this discovery,  (Stecker, de
Jager \& Salamon 1992) (hereafter  SDS) proposed that the study of the
TeV spectra of  such sources could be used  to probe the intergalactic
infrared radiation.

SDS proposed to look for  the energy dependent features indicating the
mutual annihilation of very high energy \gray s and low energy photons
of  galactic   origin  via  the  process   of  electron-positron  pair
production $\gamma + \gamma \rightarrow  e^+ + e^-$. The cross section
for this  process is  exactly determined; it  can be  calculated using
quantum electrodynamics (Breit \&  Wheeler 1934).  Thus, in principle,
if one  knows the  emission spectrum of  an extragalactic source  at a
given  redshift,  one can  determine  the  column  density of  photons
between the source and the Earth as a function of redshift.

Since  the  EGRET  discovery  of  3C279, the  infrared  background  at
wavelengths not totally dominated by galactic or zodiacal emission has
been measured by the  Cosmic Background Explorer (COBE).  In addition,
there have  been extensive observations  of IR emission  from galaxies
themselves, whose total  emission is thought to make  up the cosmic IR
background (see review  by Hauser \& Dwek 2001).  The latest extensive
observations  have been  made by  the Spitzer  satellite.  It  is thus
appropriate to use a synoptic  approach combining the very high energy
\gray ~observations  with the extragalactic IR  observations, in order
to best explore both the  \gray ~emission from blazars and the diffuse
extragalactic  IR  radiation.   Using  this approach,  one  must  take
account of {\it both} the high energy \gray\ observations and the data
from the many galaxy observations presently available.

The  MAGIC collaboration  has recently  observed the  spectrum  of the
blazar 3C279  during a  flare which occurred  on 22--23  February 2006
(Albert  et al. 2008).   The highly  luminous blazar  3C279 lies  at a
redshift  of 0.536  (Marzioni et  al.1996). To  date, it  is  the most
distant \gray\ source observed in  the sub-TeV energy range.  Thus, as
noted  by  SDS, this  source  is  potentially  highly significant  for
probing the intergalactic background radiation.

Albert et al. (2008) used their observational data to draw conclusions
regarding  the maximum  opacity  of the  Universe  to \gray  s in  the
sub-TeV energy  range.  Their conclusions  regarding the extragalactic
background  radiation would  appear  to disfavor  the  results of  the
extensive   semi-empirical  calculations   of  the   extragalactic  IR
background spectrum  given by Stecker,  Malkan \& Scully  (2006, 2007)
(hereafter SMS).   In particular,  the limit shown  in their  Figure 2
appears to be  inconsistent with one of the SMS  models that was based
primarily  on  galaxy  studies   by  the  Spitzer  infrared  satellite
telescope.

In this paper, we will reexamine both the analysis assumptions and the
conclusions presented in  the paper of Albert \etal  Using a different
analysis  technique that  we show  to be  superior to  that  of Albert
\etal, we  find that  the observations of  3C279 are  fully consistent
with  both diffuse  IR background  models  obtained by  SMS.  We  then
discuss the implications of these results regarding both the intrinsic
energy spectrum and  luminosity of the 3C279 flare  and the opacity of
the Universe to \gray s.

\section{The diffuse extragalactic IR background} 

Various calculations of the extragalactic IR background have been made
(Stecker, Puget \& Fazio 1977;  Malkan \& Stecker (1998, 2001); Totani
\&  Takeuchi (2002);  Kneiske et  al. (2004);  Primack et  al. (2005);
SMS).  Of these models, the most empirically based are those of Malkan
\& Stecker (1998, 2001), Totani \& Takeuchi (2002) and SMS.  Since the
largest uncertainty in these  calculations arises from the uncertainty
in the temporal evolution of  the star formation rate in galaxies, SMS
assumed two different evolution models, viz., a ``baseline'' model and
a ``fast evolution'' model.   These models produced similar wavelength
dependences for the spectral  energy distribution of the extragalactic
IR background,  but gave  a difference of  roughly 30-40\%  in overall
intensity.

The  empirically  based   calculations  mentioned  above  include  the
observationally based  contributions of  warm dust and  emission bands
from  polycyclic aromatic hydrocarbon  (PAH) molecules  and silicates,
which  have  been  observed  to  contribute  significantly  to  galaxy
emission in the mid-IR (e.g.,  Lagache et al. 2004).  These components
of galactic  IR emission have the  effect of partially  filling in the
``valley'' in the mid-IR spectral energy distribution between the peak
from starlight emission  and that from cold dust  emission.  The model
of  Primack et  al. (2005),  which was  based on  strictly theoretical
galaxy  spectra,  does not  take  the  warm  dust, PAH,  and  silicate
emission  components  of  mid-IR   galaxy  spectra  into  account  and
therefore exhibits  a steep mid-IR  valley that is in  direct conflict
with  solid lower  limits obtained  from galaxy  counts  obtained from
observations of  galaxies at mid-IR wavelengths (Altieri  et al. 1999;
(Elbaz  et  al2002).   This  is  clearly  shown in  Figure  2  of  the
supplemental online material of  Albert et al. (2008).  However, since
Albert et al.  (2008) considered it  to be a ``lower limit'' model, we
will discuss the Primack \etal model in our analysis.

\section{The observed spectrum of 3C279 and its derived intrinsic 
spectrum}

According  to the  MAGIC  analysis, the  2006  flare on  3C279 had  an
observed spectral index of 4.11 $\pm$ 0.68 in the energy range between
$\sim  80$ and $\sim  500$ GeV.   In their  analysis, the  MAGIC group
chose to multiply their data points by $e^{\tau(E_\gamma)}$, where the
optical  depth, $\tau(E_\gamma))$ is  chosen by  using the  results of
various  optical depth  calculations. They  then fit  simple power-law
spectra to  the resulting fluxes.  Using estimated  optical depths for
only the fast  evolution model of SMS, they  gave a best-fit power-law
spectral index  for the intrinsic  source spectrum of  $\Gamma_{s}$ of
0.49 $\pm$ 1.19 for the which  they imply is ruled out by invoking the
questionable  criterion  $\Gamma_{s}  \ge  1.5$ (However,  see,  e.g.,
(Katarzy\'{n}sky  et al.   2006;  Stecker, Baring  \& Summerlin  2007;
Resmi \& Bhattacharya 2008).

In  this paper, we  will adopt  a different  method for  analyzing the
intrinsic spectrum  of 3C279, based on  the one we first  used for the
analysis  of the  H.E.S.S. observations  of the  source 1ES0229  + 200
(Stecker \& Scully 2006). The  method is superior to the approach used
by the MAGIC collaboration where  one chooses to force a power-law fit
to the  implied {\it  deabsorbed} data points.   Instead we  assume an
intrinsic power-law  spectrum emitted by  the source over  the limited
observed energy  range that  covers less than  a decade in  energy. In
order to compare with the  observations, we multiply this power law by
an   absorption  factor  $e^{-\tau(E_{\gamma},z=0.536)}$,   where  the
optical depth, $\tau$, is calculated for a redshift z = 0.536. We then
employ a nonlinear least squares fit of our two parameter model to the
observational data.

Our  method has  several  advantages  over that  chosen  by the  MAGIC
collaboration.   To begin  with, the  redshift of  the source  and the
energy range of  the observations implies that even  the counts in the
lowest MAGIC energy bin have been  affected by at least some amount of
intergalactic  absorption.   As  a  result, multiplying  the  data  by
$e^\tau$ and  then fitting  an arbitrary power-law  as done  by Albert
\etal (2008) does not allow a proper fit to the {\it normalization} of
the spectrum. Both the true blazar luminosity and the true form of the
intrinsic  spectrum  are masked.   Furthermore,  the actual  intrinsic
spectrum  cannot be  assumed  to  have a  power-law  form {\it  after}
multiplication by an  exponential that is nonlinear in  energy at this
particular redshift (see, however,  Stecker \& Scully 2006).  Also, in
order to  properly account  for the effect  of the optical  depth, one
should directly include it in the unfolding method used to produce the
fluxes  from the  raw  photon  counts.  Since  the  MAGIC data  points
represent a  mean for an  energy bin, it  is not sufficient  to simply
multiply    that    mean   point    by    $e^\tau$.    The    function
$e^{\tau(E_\gamma)}$ is  a rapidly  changing function of  energy. This
fact must be  taken into account when computing  both the spectrum and
error bars.  The implied weighting towards a lower energy within a bin
indicates  that the appropriate  value for  $\tau(E)$ should  be lower
than that assumed for the mean energy in the bin. Thus, the values for
$\tau$ used  in the  MAGIC analysis are  too high  for all of  the EBL
models they used.

\section{Analysis}

We begin our analysis program by calculating the optical depths in the
energy range of  the observations for z = 0.536 for  both the SMS fast
evolution model and baseline  model.  For comparison, we also consider
the  optical depths taken  from Primack  \etal (2005)  and two  of the
models of Kneiske \etal (2004)  , viz., their best-fit model and their
low-IR  model. We  find  that all  five  of the  optical depth  models
considered are well fit by third order polynomials in the energy range
of interest of the form shown in equation (1).

\begin{equation}
\log   {\tau}=a_1  \log^3   E_\gamma+a_2   \log^2  E_\gamma+a_3   \log
E_\gamma+a_4
\end{equation}
Table 1  summarizes the fit  parameters to the five  models considered
here.   Figure 1  shows the  excellent  agreement of  the third  order
polynomial fits to optical depths for the five models.
\begin{deluxetable}{lllll}
\tablecaption{Parameters  Used  in Equation  (1)  for Opacity  Models}
\tablewidth{0pt}    \tablehead{\colhead{Model}    &    \colhead{$a_1$}
&\colhead{$a_2$}  &\colhead{$a_3$} & \colhead{$a_4$}}  \startdata Fast
Evolution &  0.0999257 & -1.26786 &  5.36642 & -6.21713  \\ Baseline &
0.103179  &  -1.28947 &  5.41409  & -6.36993  \\  Kneiske  Best Fit  &
-0.477755  &  2.84965 &  -3.77753  & -0.353419  \\  Kneiske  Low IR  &
-0.334462 &  1.75885 &  -1.19612 & -2.30134  \\ Primack &  -0.549224 &
3.43815 & -5.63807 & 1.57098 \\ \enddata
\end{deluxetable}

\begin{figure}[th] 
\begin{center}
\includegraphics[scale=1.0]{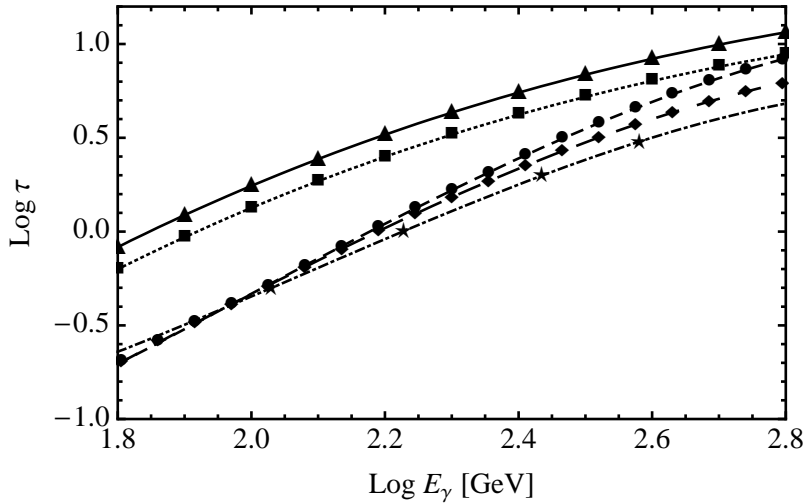}
\end{center}
\caption{Polynomial fits to the five optical depth models discussed in
the text.   The solid line  and dotted line  are fits to the  SMS fast
evolution (triangles) and baseline (filled boxes) models respectively.
The  short-dashed and  long-dashed lines  are polynomial  fits  to the
best-fit Kneiske \etal model (filled circles) and low-IR Kneiske \etal
model (diamonds) respectively.  The  dot-dashed line is the polynomial
fit to the Primack \etal optical depth model (stars).}
\label{taufit}
\end{figure}

\begin{figure}[t] 
\begin{center}
\includegraphics[scale=1.0]{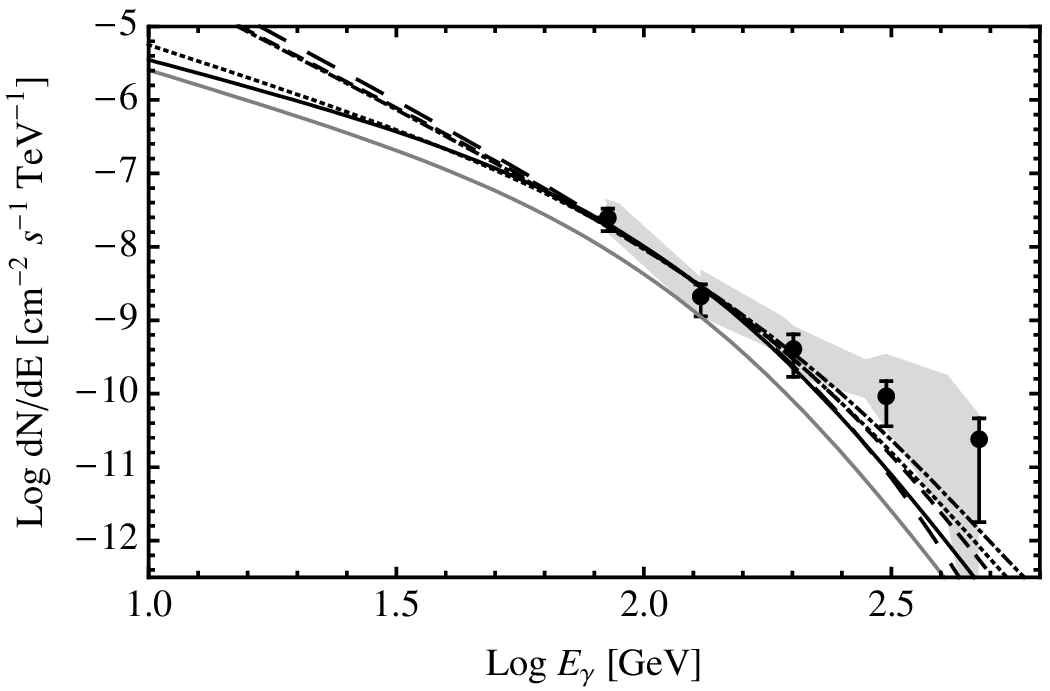}
\end{center}
\caption{Plot of the best-fit curves with $\Gamma_{s}$ and $K$ as free
parameters for the SMS  fast evolution (solid), SMS baseline (dotted),
Kneiske   \etal   best-fit   (short-dashed),  Kneiske   \etal   low-IR
(long-dashed), and Primack \etal (dash-dotted) models shown along with
the  observational data from  Albert \etal  (2008). The  shaded region
represents  the  combined systematic  and  statistical  error for  the
observational data. The gray curve shows the EGRET flare power-law fit
extended  to  higher   energies  multiplied  by  $e^{-\tau(E)}$.   The
respective $\chi^2$  values are  given in the  third number  column in
Table 2.}
\label{bestfit}
\end{figure}

Because  of the  nonlinear  nature  of the  energy  dependence of  the
optical depth, we do not expect that the observed spectrum will have a
power-law  form.  Thus, we  fit  to spectra  that  are  assumed to  be
power-law   intrinsic   source   spectra  ($K   {E_\gamma}^{-\Gamma}$)
multiplied  by $e^{-\tau}$  as prescribed  by equation  (1)  using the
various optical depth  models indicated in Table 1  to take account of
intergalactic  absorption.   We  then employ  the  Levenberg-Marquardt
nonlinear least squares method  to fit the resulting nonlinear spectra
to  the observed  data.   We choose  to  fit to  the  data using  only
statistical errors.

We  first do  a $\chi^2$  a fit  with two  free parameters,  viz., the
normalization coefficient, $K$ and the intrinsic spectral index of the
source $\Gamma_{s}$. The best-fit  spectral curves for the various EBL
models considered are shown in Figure \ref{bestfit}. The corresponding
best-fit values  $K$ and  $\Gamma_{s}$ are given  in Table 2.  In this
table,  in the  first column,  the  $K$ values  are given  for $E$  in
GeV. The second column gives  the best-fit spectral indeces for all of
the  models.  The third  column  gives  the  $\chi^2$ values  for  the
two-parameter  best fit. It  can be  seen from  this column  that 
although the best-fit spectral indeces are significantly different for
the various  EBL models, the  best fit $\chi^2$ values  are comparable
for all of the models.

\begin{deluxetable}{lllll}
\tablecaption{Best-fit Spectral Parameters for EBL Models with $\chi^2$ Values}
\tablewidth{0pt}              
\tablehead{\colhead{Model}  & \colhead{$K$}
& \colhead{$\Gamma_s$} & \colhead{$\chi^2$} & 
\colhead{$\chi^2$ for $\Gamma_{s}=2.02 $ }}
\startdata Fast Evolution & 2.15  & 1.78 & 1.59 & 1.20  \\ 
Baseline &  8.75 & 2.19 & 1.42 & 1.07 \\ 
Kneiske Low IR & 128 & 3.46 & 1.59 & 1.28 \\ 
Kneiske Best Fit & 283 & 3.62 & 1.55 & 1.42 \\ 
Primack & 141 & 3.49 & 1.54 & 1.25 \\ 
\enddata
\end{deluxetable}

The Levenberg-Marquardt method yields substantially different best-fit
spectral indices for  our modeled optical depths and  those of Kneiske
\etal (2004)  and Primack \etal  (2005).  However, these fits  are not
particularly  unique in  their goodness  of fit. We performed
a $\chi^2$ analysis on all of the models, allowing both
$\Gamma_{s}$ and  $K$  to be free parameters  in the  fit. As can be seen
from Table 2, {\it  the form of the  intrinsic spectrum of the  3C279 flare is
undetermined}   by  the  five   data  bins   obtained  by   the  MAGIC
collaboration,  even when  one  only considers  the large  statistical
errors involved and neglects the significant systematic errors. This is
because the lower energy points with the smaller
error bars are weighted more highly than the highest energy point. 

Since  the minimum $\chi{^2}$ is almost independent of
choice  of spectral  index, the  two-parameter fitting  routine mainly
tries  to  move  the  curve  up  and  down  ({\it  i.e.}  adjusts  the
normalization) to  best fit the  observational data.  This  feature is
lacking in the MAGIC analysis of the spectrum of 3C279 since they have
factored in the absorption effect {\it prior} to making their fits.

\begin{figure}[t] 
\begin{center}
\includegraphics[scale=1.0]{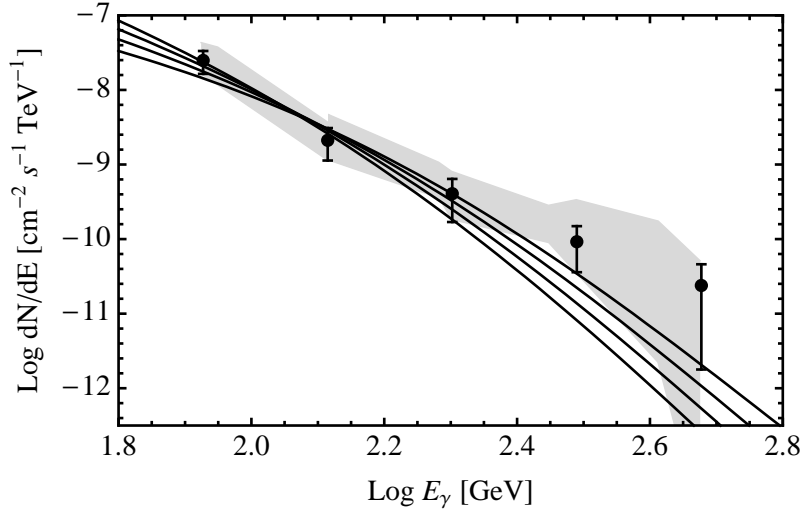}
\end{center}
\caption{Plot of the  best fit curves obtained using  the SMS baseline
model for fixed spectral indices of 1.5, 2.0, 2.5, and 3.0. The shaded
region represents  the combined  systematic and statistical  error for
the observational data.}
\label{multispec}
\end{figure}

To further illustrate this point of ambiguity in the spectral index of
the source, Figure \ref{multispec} shows the best fit obtained for the
SMS baseline model to the observational data for the range of spectral
indices indicated. Spectra  with the higher values of $\Gamma_s$
provide a  good fit to the  three lower energy data  points which have
the smallest  error bars.  However,  they miss the higher  energy data
points.  Those with  lower values of $\Gamma_s$ miss  the lower energy
data point but  pass through the four higher energy  data points. 
all of these  fits have  almost identical  reduced $\chi{^2}$
values of $\sim$ 1.1. Choosing  a range  of $\Gamma_s$  between 1.5  
and 3 as shown in Figure \ref{multispec}, and varying the normalization
constant,  $K$ to find the best fit in each case, it is found that  
all of these spectral fits have  
almost identical reduced $\chi{^2}$ values of $\sim$ 1.1.
As a result, all of them are equally probable. Other EBL models
exhibit the same degeneracy in $\chi^2$ over a range of spectral
indeces.

\section{Hypothetical Consideration of the Luminosity of the 
3C279 Flare Observed by MAGIC}

As discussed  above, our statistical analysis shows  that the spectral
index of  the flare is not  well determined by the  MAGIC data.  Given
this uncertainty,  we can conditionally compare the  luminosity of  the MAGIC
flare with  that of  an earlier flare  observed by EGRET by assuming the
same intrinsic spectral index as the EGRET flare. The 1991 EGRET flare
had  an observed  spectral  index  of $\Gamma_s  =  2.02$ at  energies
between 50 MeV  and 10 GeV (Hartman et al.  1992) where EBL absorption
is negligible. 

We note that this choice of spectral index gives good
$\chi^2$ fits for all the EBL models considered here.
Assuming the value for $\Gamma_s$ of  2.02, we obtain the best fits to
the  MAGIC data using  the Levenberg-Marquardt  method for  the single
free parameter, $K$.  The $\chi^2$ values obtained for these 
best fits are shown in the last column of Table 2.
The resulting curves  for the fits of all five models are 
illustrated in figure \ref{bestfit}.

If  we  then choose  the  form  of $\tau(E)$  given  by  the SMS  fast
evolution model as input to equation (1), by extending the $\Gamma_s =
2.02$ power-law  fit from Hartman et  al.  to an energy  of $\sim$ 500
GeV, we  find a value for  $K$ that is  $\sim$ 3 times higher  for the
MAGIC  flare   than  for  the   flare  observed  by  EGRET   15  years
earlier. Thus, we  obtain a reasonable estimate for the  luminosity 
of the MAGIC flare  that is comparable  to that  of the  earlier flare  
observed by EGRET.

\begin{figure}[t] 
\begin{center}
\includegraphics[scale=1.0]{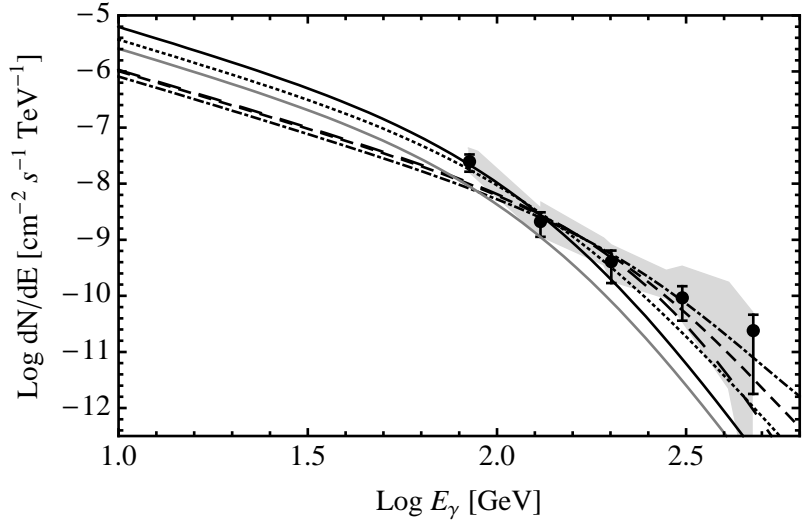}
\end{center}
\caption{Plot  of the  best  fit  curves with $\Gamma_{s}=2.02$ 
for  the  SMS fast  evolution
(solid), SMS baseline (dotted), Kneiske \etal best-fit (short-dashed),
Kneiske  \etal low-IR (long-dashed),  and Primack  \etal (dash-dotted)
models for  a fixed  spectral index of  $\Gamma_s = 2.02$  shown along
with  the observational  data  from Albert  \etal  (2008). The  shaded
region represents  the combined  systematic and statistical  error for
the observational data. The gray curve shows the EGRET flare power-law
fit extended to higher energies multiplied by $e^{-\tau(E)}$. The respective
$\chi^2$ values are given in the last column of Table 2.}
\label{202}
\end{figure}

\section{Conclusions and Discussion}

We  conclude  that  the observational data of  3C279 from
the MAGIC collaboration do not significantly constrain the
intergalactic low energy photon spectra, nor do they indicate
that the Universe is more transparent to high energy \gray s than
previously thought or obtained by the SMS models.
Including  the effects  of  the systematic  errors,  which only  
significantly affect the two highest energy  data points,  
would  only strengthen  our conclusions. 

We show here, as well as in our analysis of 1ES0229 (Stecker \& Scully
2008),   that  the  SMS   models  produce   reasonable  fits   to  the
observational data.   We further demonstrate that other  models in the
literature that  give lower transparencies  do not fit the  3C279 data
better  than the SMS  models do.   This is  because the  magnitude and
shape of the intrinsic spectrum of the flare are not well determined 
by the MAGIC data, as we have shown.

The spectral  energy distribution  of the intergalactic IR radiation 
is well  constrained by
astronomical data, as discussed in detail in SMS. However, the form of
the  intrinsic  source  spectrum  of  the  \gray  ~flare  is  not  well
constrained.  If  one should wish to  speculate on producing  a fit to
the  MAGIC data  to go  through  {\it all}  of the  points, one  would
require an intrinsic flare spectrum which flattens at higher energies.
Such a spectrum may  arise either from relativistic shock acceleration
(Stecker,  Baring   \&  Summerlin  2007; Resmi \& Bhattacharya 2008)  or   
from  intrinsic  source
absorption (Liu \& Bai 2006; Liu, Bai \& Ma 2008; Sitarek \& Bednarek 2008).   
However, given  both the  statistical and  the systematic
errors of  the MAGIC data,  particularly those for the  highest energy
bin, the invocation of such processes is unnecessary.

If we apply our analysis technique
to the MAGIC data and assume an intrinsic power-law spectrum with
index 2.02 as observed for the earlier EGRET flare at lower energies
where absorption is negligible, we find that the luminosity of the
MAGIC flare was similar to that of the earlier EGRET flare. In fact,
the MAGIC flare was $\sim$3 times brighter. This luminosity estimate
could not be obtained using the analysis adopted by the MAGIC group,
since {\it all} of the data points that they started from were
affected by intergalactic absorption.

One would  need  to have  additional
information   concerning  the  intrinsic   source  spectrum   such  as
additional observations at lower \gray\ energies that are unaffected by
pair-production absorption in order to better constrain the intergalactic
IR radiation through
an  analysis of  the resulting  optical depth at the  higher energies
where absorption  is more significant. 
We note that the ideal situation for exploring the exact amount
of intergalactic absorption would be to have simultaneously
and with overlapping observational energy ranges (1) observations of a
strong flare with large photon counts at lower energies unaffected by
intergalactic absorption, as can be obtained by FGST (Fermi Gamma-ray
Space Telescope, nee GLAST), and (2) ground based sub-TeV observations 
of such a strong flare with larger photon counts.

\acknowledgements
We wish to  thank Rudolf Bock for  sending us a list of  data on the
spectrum of 3C279 observed by MAGIC. STS gratefully acknowledges
partial support from the Thomas F. \& Kate Miller Jeffress Memorial Trust 
grant no. J-805.

\end{document}